\documentclass[pss]{wiley2sp} 
\usepackage{amsmath}
\usepackage{bm}             

\begin{document}

\title{Spin Transport in Armchair Silicene Nanoribbon on the Substrate: The Role of Charged Impurity}

\author{%
  Shoeib Babaee Touski\textsuperscript{\Ast}}

\mail{
  \textsf{touski@hut.ac.ir}}

\institute{%
  Department of Electrical Engineering, Hamedan University of Technology, Hamedan 65155, Iran}


\abstract{\bf
In this work, electrical and spin properties of armchair silicene nanoribbon (ASiNR) in the presence of charged impurity is studied. The non-equilibrium Green's function along with multi-orbital tight-binding is applied to obtain transmission probability. different type of spin transmission probability in the ASiNR on a substrate is investigated. The charged impurities are located in the underlying substrate. Spin-flip along the channel is calculated by using spin transmission probability. Spin diffusion length in ASiNR for differently charged impurities is obtained and compared with the mean free path.}

\maketitle   

\section{Introduction}
Two-dimensional (2D) silicene with a honeycomb arrangement identical to graphene have attracted high attention thanks to their unique electronic and spintronic properties \cite{kara2012review,houssa2015silicene,lew2016silicene}. Two-dimensional atomic layers are characterized for application in the future electronic devices owing to their high charge mobility, ultra-thin body, and other excellent properties. Opposite to graphene, silicene possesses a buckled structure with a relative large intrinsic spin-orbit coupling (SOC), opening a band gap at Dirac points and can present the quantum spin Hall effect (QSHE) \cite{ezawa2012valley}. Silicene inherently is compatible with silicon semiconductor technology that potentially can be a good candidate in spintronic devices \cite{liu2011quantum,liu2011low}.

Silicene nanoribbons have significant electronic properties applicable to electric and spintronic devices \cite{tsai2013gated}.
silicene has two different edge terminations, i.e., zigzag and armchair. In a zigzag-edge nanoribbon, electronic and magnetic properties are dominated by edge states.
Zigzag silicene nanoribbons (ZSiNRs) has been studied for application in spintronic devices. Half-metallicity can be realized by transverse electric filed and asymmetric edge modification \cite{ding2013electronic,yang2014transport,zhang2014bipolar,deng2014edge}. Furthermore, Spin-filter has been reported based on its half-metallicity. A giant magneto-resistance is achieved in ZSiNRs by switching between two different magnetic states or symmetry-dependent transport property \cite{xu2012giant,kang2012symmetry}. 
Half-metallicity can be obtained in hydrogenated silicene and an out of plane gate voltage \cite{zhang2012first}. Electric field or an exchange field can control quantum spin Hall effects (QSHE), quantum anomalous Hall effects (QAHE) and quantum valley Hall effects (QVHE) in pristine silicene\cite{ezawa2012valley}.

Silver is reported as the common substrate for silicene \cite{aufray2010graphene,vogt2012silicene}.
Orbital bonding Si atoms to Ag substrate vanishes the Dirac cone in the epitaxial silicene on Ag substrate \cite{vogt2012silicene,guo2013absence}. 
The substrate influences on the supporting silicene so that it significantly changes the electronic properties of silicene \cite{lin2013substrate}. 
The band gap of silicene on the substrates is enlarged compared to suspended silicene\cite{liu2013silicene}. 
Silicene field effect transistor has been reported on the SiO$_2$ substrate that operates at room-temperature and ambipolar current indicates Dirac band structure \cite{tao2015silicene}.  Al$_2$O$_3$ dielectric substrate also have a weak effect on silicene band structure opposite to Ag substrates \cite{chen2016designing,aufray2010graphene}.
Silicene has been grown on some substrates, such as Ir, BN and SiC \cite{feng2012evidence,meng2013buckled,kaloni2013quasi}. Some substrates such as  MgX$_2$ (X=Cl, Br and I), GaS and BN substrates can approximately preserve Dirac cone\cite{kaloni2013quasi,kokott2013silicene,liu2013silicene,ding2013electronic,zhu2014structural}.

Substrates affect electrical transport in two main types of disorder: charged impurities and surface roughness.  Ripples reported in several experiments that are a major source of disorder \cite{meyer2007structure,ishigami2007atomic,kim2008graphene}. Our previous report, S.B. Touski et al, \cite{touski2013substrate}, indicates extrinsic ripple from substrate play a dominant role in the electrical transport in a graphene nanoribbon. On the other hand, some experiments approve that charged impurities are the dominant scatterers and high-K gate dielectric can increase mobility with the screening of charged impurity \cite{ponomarenko2009effect,ma2014charge}. 
The disordered potential will be the main scattering mechanism based upon experimental \cite{sarma2011electronic,bolotin2008ultrahigh}.
 
In this paper, we explore electric and spin transport in an armchair silicene nanoribbon under charged impurity. 
In our previous works, spin transport at the presence of surface roughness in graphene nanoribbon \cite{chaghazardi2016spin} and MoS2 and WS2 \cite{touski17} were studied. It was shown surface roughness has a high impact on the spin-flip for both materials. Here, we investigate the effect of substrate's charged impurity on the spin transport.

\section{Approach}

Hamiltonian of the silicene in multi-orbital consideration can be expressed as:
\begin{equation}
H=\sum_{i;ls} \epsilon_{i;ls} c_{i;ls}^\dagger c_{i;ls} + \sum_{\left\langle i,j \right\rangle;l,m} V_{i,j;l,m} c_{i;l}^\dagger c_{j;m}+H_{SO},
\label{Eq:ham}
\end{equation}
where $i,j$ are the atomic position, $l$ and $m$ run over the atomic orbitals, $c_{i;l}^\dagger (c_{i;l})$ creates (annihilates) an electron at orbtial $l$ of site $i$, $\epsilon_{i;l}$ refers to on-site crystal fields of orbital $l$ and $V_{i,j;l,m}$ are hopping parameters, where $\langle ij\rangle$ runs over first nearest neighbor sites. The hopping parameter is taken from Ref.\cite{liu2011low} where $V_{ss\sigma}=-1.93$, $V_{sp\sigma}=2.54$, $V_{pp\sigma}=4.47$, $V_{pp\pi}=-1.12$, $\epsilon_s=0$ and $\epsilon_p=-7.03$. The last term $H_{SO}$ indicates spin-orbit Hamiltonian. \begin{equation}\label{Eq:HSO}
H_\mathrm{SO}= \sum_{i;l,m}\frac{\lambda}{\hbar} L_{i;l}\cdot S_{i;m},
\end{equation}
where $\lambda$ is the intra-atomic SOC constant that is taken $34meV$ \cite{liu2011low}. $L$ is the angular momentum operator for atomic orbitals, and $S$ is the spin operator.

For modeling charged impurity disorders, a superposition of Gaussian potential fluctuations is added to the Hamiltonian as a diagonal term.
The electric potential of all charged impurities can be expressed by the sum of potentials due to the individual charged impurities with using the superposition principle. Charged impurity in the substrate is screened and can be modeled by a Gaussian potential as\cite{rycerz07,lewenkopf08,djavid2014}:
\begin{equation}
U_{imp}(r)=\sum_{n=1}^{N_{imp}}  U_n exp\left(-\frac{|r-R_n|^2}{2\xi^2}\right),
\end{equation}
where $N_{imp}$ is number of impurity sites and  Let $n_{imp} = N_{imp} /N$ denotes the density of scatterers. $U_n$ shows potential amplitude that is randomly distributed in the interval $[-\delta U,\delta U]$ and $\xi$ denotes to potential range. $N_{imp}$ Gaussian potential with potential range $\xi$ is Uniformly distributed at random sites $R_n$. $U_{imp}$ is added to Hamiltonian as on-cite potential.

The non-equilibrium Green's function (NEGF) formalism \cite{pourfath14non} is used to investigate spin transport in armchair silicene nanoribbons. The retarded $\underline{G}^r$ and advanced $\underline{G}^a$ Green's functions is calculated by:
\begin{equation}
\begin{split}
\underline{G}^r(E)&=[(E+i\delta)\underline{I}-\underline{H}-\underline{U_{imp}}-\underline{\Sigma}^L - \underline{\Sigma}^R]^{-1},\\
\underline{G}^a(E)&=[(E-i\delta)\underline{I}-\underline{H}-\underline{U_{imp}}-\underline{\Sigma}^L - \underline{\Sigma}^R]^{-1},
\end{split}
\end{equation}
where $E$ is the energy, $\underline{I}$ is the identity matrix and $\delta$ is a phenomenological broadening (10$^{-5}$eV), and $\Sigma^{L,R}$ is the self-energy of the left and right contacts
\begin{equation}
\Sigma^{L,R} = \tau^{L,R}g^{L,R}
\left(\tau^{L,R}\right)^\dagger,
\label{Eq:sigma}
\end{equation}
where $g^{L,R}$ is the surface Green's function of the contacts, given by
\begin{equation}\label{Eq:g}
\underline{g}^{L,R} = \left[E\underline{I}-\underline{H}^{L,R}-\underline{h}_{c}^{L,R} \underline{g}^{L,R}\left(\underline{\tau}_{c}^{L,R}\right)^\dagger\right]^{-1}\ ,
\end{equation}
that $\underline{H}^{L,R}$ is the Hamiltonian of the unit cell of the right or left contact in real space representation, $\underline{h}_c^{L,R}$ is the coupling between two neighboring unit cells in the considered contacts, and $\underline{\tau}_{c}^{L,R}$ is the coupling between the channel and the contacts. Underlined quantities stand for matrices that include both spins.
For the calculation of the contact self-energies, the surface Green's function of the contacts is iteratively solved, employing a highly convergent scheme \cite{sancho1985highly}.

The spin-resolved transmission probability can be written as:  
\begin{equation} 
\begin{split}
T_{\sigma\sigma'}(E) &= {\rm Tr} \left[ \Gamma^L_{\sigma}G^r_{\sigma\sigma'}\Gamma^R_{\sigma'}G^a_{\sigma'\sigma} \right],
~~~~\sigma,~\sigma'=\uparrow,\downarrow,
\end{split}
\label{Eq:fourTE}
\end{equation}
where $\Gamma^{L,R}_\sigma = i\left(\Sigma^{L,R}_\sigma -\left( \Sigma^{L,R}_\sigma{}\right)^\dagger\right)$  
describes the broadening of the two semi-infinite leads.
$T_{\uparrow\uparrow}(E)$ and $T_{\downarrow\downarrow}(E)$ represent parallel spin transmission, and $T_{\uparrow\downarrow}(E)$ and $T_{\downarrow\uparrow}(E)$ antiparallel spin-flip transmission. Detailed of the spin NEGF can be found in our previous works \cite{chaghazardi2016spin,touski17}.

\section{Results}

\begin{figure}[t]
	\centering
	\includegraphics[width=1.0\linewidth]{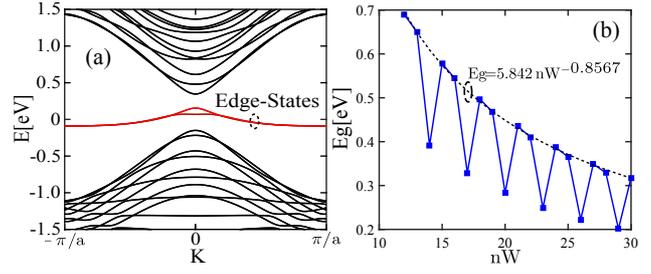}
	\caption{(a) Band structure of a ASiNR with $\mathrm{nW=18}$. Edge-states is indicated in the middle of band-gap. (b) Variation of band-gap as a function of nanoribbon width. }
	\label{fig:fig1}
\end{figure} 

Hamiltonian for ASiNR with using Eq. \ref{Eq:ham} and \ref{Eq:HSO} has been obtained and band structure is calculated. Band structure for an ASiNR with $\mathrm{nW=18}$ (nW accounts atoms at the width of nanoribbon) is plotted in the Fig .\ref{fig:fig1}(a). One can see the two bands in the band gap that is contributed to edge states. Band-gap is calculated from band structure without edge states consideration, see Fig. \ref{fig:fig1}(a). The amount of band gap is approximately compatible with Ref. \cite{cahangirov10}. Band-gap totally decreases with increasing nanoribbon width, however, nanoribbons with $nW=3n+2$ have lower band-gap to others. Bandgap can be tuned from $\mathrm{E_g=0.7eV}$ to $\mathrm{0.2eV}$.

\begin{figure}
	\centering
	\includegraphics[width=0.999\linewidth]{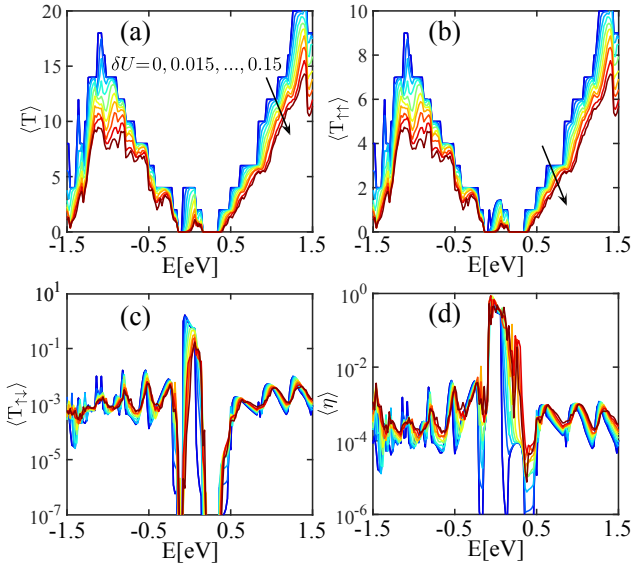}
	\caption{a) Total transmission probability, b) up- to up-spin transmission, c) up- to down-spin transmission, and d) spin efficiency as a function of energy for different $\delta U$. $\mathrm{nW=18}$, $\mathrm{L=20nm}$, $\mathrm{n_{imp}=0.01}$, and  $\mathrm{\xi=3a_{Si-Si}}$. }
	\label{fig:fig2}
\end{figure}

Electrical potential from charged impurity is added to nanoribbon's Hamiltonian and transmission probability is calculated with using the NEGF method. 64 samples with randomly charged impurities at random position are created then the mean average indicates the behavior of the ASiNR.
$\delta U$ is changed from 0 to $\mathrm{0.15eV}$ and the results are plotted in Fig. \ref{fig:fig2}. One can see total transmission decreases due to increasing of scattering from charged ions. 
Transmission from up-spin to up-spin ($T_{\uparrow\uparrow}$) follows approximately total transmission, Only transmission at edge-states decreases. $T_{\downarrow\downarrow}$ is completely similar to $T_{\uparrow\uparrow}$ and is not reported here. On the other hand, transmission from up- to down-spin ($T_{\uparrow\uparrow}$) at these edge-states is comparable with $T_{\uparrow\uparrow}$. $T_{\uparrow\downarrow}$ decreases for some energies, however, increases for some others. $T_{\uparrow\downarrow}$ is contributed to two mechanisms. First, increasing spin-flip increases $T_{\uparrow\downarrow}$, then, scattering from charged ions declines this transmission. To separate spin-flip from scattering, spin efficiency can be defined as:
\begin{equation}
\eta(E)=\frac{T_{\uparrow\downarrow}}{T_{\uparrow\downarrow}+T_{\uparrow\uparrow}}.
\end{equation}
Wave-like of spin efficiency is clear from Fig. \ref{fig:fig2}(d). Spin efficiency behaves like to $T_{\uparrow\downarrow}$, increases for some energies and decreases for others. $\eta$ separates scattering and presents spin-flip. Because scattering from CI is small, $\eta$ and $T_{\uparrow\downarrow}$ behave similarly. We found Spin-flip increases for first sub-bands in conduction band third sub-band, whereas, decreases for second and fourth sub-band. Spin-flip increases for odd sub-bands and decreases for even ones.

\begin{figure}
	\centering
	\includegraphics[width=1.0\linewidth]{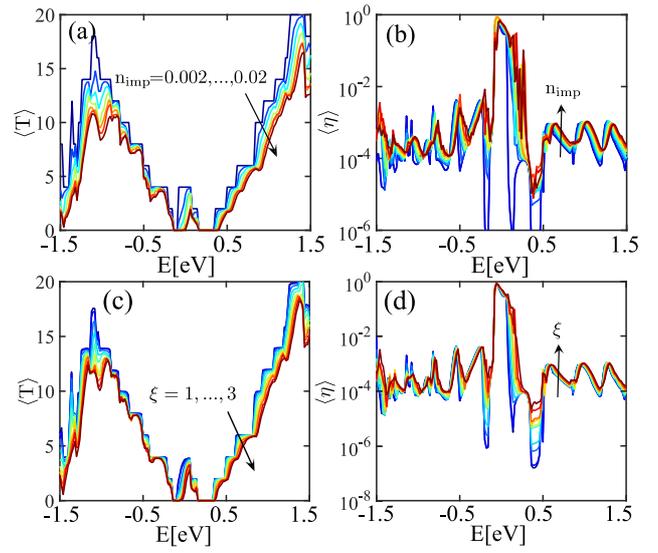}
	\caption{(a) Total transmission probability and (b) spin efficiency as a function of energy for different charged impurity density. (c) Total transmission and (b) spin efficiency versus energy for different charged impurity density.$\mathrm{nW=18}$, $\mathrm{L=20nm}$, $\mathrm{\xi=3a_{Si-Si}}$ and  $\mathrm{\delta U=0.1eV}$. }
	\label{fig:fig3}
\end{figure}

The density of charged impurity is another important parameter that indicates the quality of the underlying substrate. CI density is swept from $n_{imp}=0.002$ to $n_{imp}=0.02$ \cite{rycerz07,lewenkopf08} and resulted is shown in the Fig. \ref{fig:fig3}(a) and (b). As one can see, transmission probability decreases with increasing the impurity density due to scattering rate increment with the increasing number of scatterers. The effects of density on the transmission and spin efficiency follow the behavior of charge amplitude. One can observe a linear dependency between charged amplitude and density for charge and spin transports.

The underlying substrate and supporting dielectric screens potential of charged impurities. Different substrates and dielectrics enforce different potential ranges. The potential range is changed from $a_{Si-Si}$ to $3a_{Si-Si}$ in the long-range potential where $a_{Si-Si}$ is the distance between two neighbors Si atoms and the results are plotted in Fig. \ref{fig:fig3}(c) and (d). one can observe the effects of $\xi$ is like to $U_{imp}$ and $n_{imp}$ and lower than them.

\begin{figure*}[htb]
	\centering
	\includegraphics[width=\textwidth]{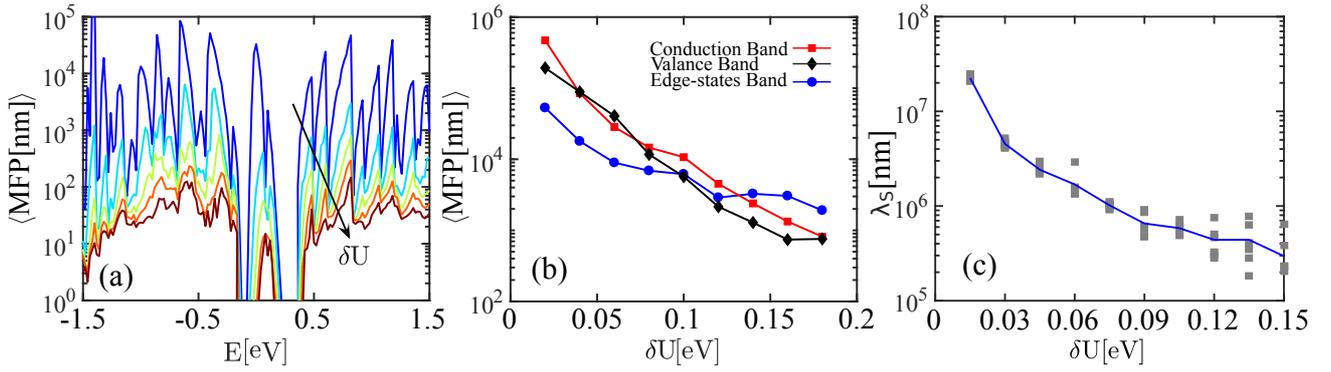}
	\caption{(a) Mean free path as a function of energy at different charge amplitudes. (b) Mean free path as a function of $\delta U$ for conduction, valance and edge bands. (c) Spin diffusion length versus  charge amplitude for conduction band. $\mathrm{nW=18}$,$\mathrm{n_{imp}=0.01}$ and $\mathrm{\xi=3a_{Si-Si}}$.}
	\label{fig:fig4}
\end{figure*}

The channel length of ASiNR is varied and its spin transport is studied. Total transmission decreases as channel length increases. Mean free path (MFP) in ASiNR can be obtained with \cite{yazdanpanah12}:
\begin{equation}
\lambda(E)=\frac{L}{N/T(E)-1}
\label{eq:mfp}
\end{equation}
Where N is the number of sub-band at each energy. Total transmission is investigated versus channel length and MFP is extracted by fitting Eq. \ref{eq:mfp}. MFP as a function of energy is plotted in Fig. \ref{fig:fig4}. In the edge of each sub-band, Density of state (DOS) increases and mean free path decreases due to increasing of DOS. MFP for conduction, valance and edge bands are plotted as a function of $\delta U$ in Fig. \ref{fig:fig4} (b). MFP for conduction and valance bands behave same and approximately decreases two orders of magnitude for decreasing $\delta U$ from $\mathrm{0.02}$ to $\mathrm{0.2eV}$. However, MFP versus $\delta U$ in the edge states decreases steeply at the beginning and smoothly declines. MFP is longer than 100nm for every charged impurity in the substrate. Nowadays technology is smaller than $\mathrm{100nm}$ that means electron transmits in nanoscale as ballistic.

Electron's spin flips along the channel with the contribution of spin-orbit coupling and scattering from charged impurities. An electron injected in the channel with up-spin and reaches to right channel with up-spin with $T_{\uparrow\uparrow}$ and with down-spin with $T_{\uparrow\downarrow}$ probability.  
Polarization along the channel can be defined as:
\begin{equation}
P(E)=\frac{T_{\uparrow\uparrow}-T_{\uparrow\downarrow}}{T_{\uparrow\uparrow}+T_{\uparrow\downarrow}}
\label{Eq:pol}
\end{equation}
Where polarization at the beginning of the channel is one and decreases along the channel with increasing $T_{\uparrow\downarrow}$.

Transmission probabilities for ASiNR with different channel length in the presence of charged impurity is calculated. Polarization is obtained with Using Eq. \ref{Eq:pol}. Polarization decays with length as: $P(L)=P_0 \exp(-L/\lambda_s)$ where $\lambda_s$ is spin diffusion length (SDL) \cite{chaghazardi2016spin}.  Polarization is studied as a function of length, then SDL is extracted with an exponential fitting. SDL for electron in conduction band versus $\delta U$ is plotted in the Fig .\ref{fig:fig4}(c). SDL is in the order of $\mathrm{10^6 nm=1mm}$ that is two or three orders of magnitudes longer than MFP. This means an electron experiences 100 to 1000 scatterings before spin-flip. SDL approximately is 100 times longer for small charge amplitudes however it reaches to 1000 times for large charge amplitudes. MFP is more sensitive to charge impurity relative to SDL. The obtained SDL is much longer than the SDL reported by others. 
SP Dash, et al, \cite{dash2009} reports $\mathrm{\lambda_s=230 nm}$ for electrons and $\mathrm{\lambda_s=310 nm}$ for holes. B. Bishnoi et al \cite{bishnoi2013spin} with using Monte-Carlo reported $\mathrm{\lambda_S=500 nm}$ and D’yakonov-Perel (DP) as the main relaxation mechanism. Here, there is a correlation between SDL and MFP, however, SDL completely doesn't follow MFP. Our results indicate both D’yakonov-Perel (DP) and Elliott-Yafet
(EY) contribute to spin relaxation in silicene.  The effects of other scattering mechanisms such as surface roughness and out of plane phonon should be investigated to clarify the main source of spin relaxation in silicene.

\section{Conclusion}
We report the effects of charged impurities on both electrical and spin properties of ASiNR. Charged impurity decreases total transmission and limits electrical transport. On the other hand, charged impurities enhance spin-flip and the transmissions with spin-flip. Mean free Path is obtained for differently charged impurities that are longer than present transistor channels. Spin diffusion length is calculated that is two or three orders of magnitudes longer than MFP.  


\providecommand{\WileyBibTextsc}{}
\let\textsc\WileyBibTextsc
\providecommand{\othercit}{}
\providecommand{\jr}[1]{#1}
\providecommand{\etal}{~et~al.}



\end{document}